\documentclass[onecolumn]{aa_mod}

\usepackage{graphicx}
\usepackage{txfonts}
\begin{document}

   \title{Astrocladistics: a phylogenetic analysis of galaxy evolution
}

\subtitle{I. Character evolutions and galaxy histories}

                      \titlerunning{Astrocladistics: a phylogenetic analysis of galaxy evolution I} 
                      \authorrunning{Fraix-Burnet et al.} 

   \author{Didier Fraix-Burnet \\ \textit{Laboratoire d'Astrophysique de Grenoble, France }
          \and \\  \vskip 0.1 true cm
          Philippe Choler \\ \textit{Laboratoire d'\'Ecologie Alpine, Grenoble, France}
          \and \\  \vskip 0.1 true cm
          Emmanuel J.P. Douzery \\ \textit{Laboratoire de Pal\'eontologie, Phylog\'enie et Pal\'eobiologie, 
          \\ Institut des Sciences de l'\'Evolution de Montpellier, France}
          \and \\ \vskip 0.1 true cm
          Anne Verhamme \\ \textit{Laboratoire d'Astrophysique de Grenoble, France}
          }
\institute{
             \footnotetext{Anne Verhamme is now at Geneva Observatory, Switzerland. First author's address: Laboratoire d'Astrophysique de Grenoble, BP 53, F-38041 Grenoble cedex 9, France, email: fraix@obs.ujf-grenoble.fr. Supplementary material (numerical tables and character projections) is available on http://hal.ccsd.cnrs.fr/aut/fraix-burnet or http://www-laog.obs.ujf-grenoble.fr/public/fraix.}
 }

\date{}

   \abstract{ This series of papers is intended to present astrocladistics in some detail and evaluate this methodology in reconstructing phylogenies of galaxies. Being based on the evolution of all the characters describing galaxies, it is an objective way of understanding galaxy diversity through evolutionary relationships. In this first paper, we present the basic steps of a cladistic analysis and show both theoretically and practically that it can be applied to galaxies. For illustration, we use a sample of 50 simulated galaxies taken from the GALICS database, which are described by 91 observables (dynamics, masses and luminosities). These 50 simulated galaxies are indeed 10 different galaxies taken at 5 cosmological epochs, and they are free of merger events. The astrocladistic analysis easily reconstructs the true chronology of evolution relationships within this sample. It also demonstrates that burst characters are not relevant for galaxy evolution as a whole. A companion paper is devoted to the formalization of the concepts of formation and diversification in galaxy evolution. 
   
   }

   \maketitle
\newpage


\section{Introduction}
\label{intro}

Classification is a very human act to release memory and understand relationships between different kinds of objects. Systematics is the science of classification of living organisms. It teaches us that there are three ways of comparing complex objects generated during the course of evolution: appearance, global similarity and common history. 

The first approach was used by the Greeks (Aristotle) and until the 18th century. It was based on the selection of one or very few obvious patterns of living organisms. This selection was necessarily subjective and with the discovery of more and more specimens during the Middle Ages, numerous classifications appeared. They became rapidly awkward, incompatible with one another, and unsatisfactory in the representation of the origin of the observed 
diversity (see e.g. Knapp~\cite{knapp}).

In comparison, extragalactic astrophysics is a very young science beginning when Hubble discovered the true nature of galaxies in 1922 (Hubble~\cite{h22}). He rapidly felt the need to classify these new objects using the only information he had, i.e. morphology (Hubble~\cite{h26} and Fig.~\ref{hubbleclad} below). In an attempt to understand the relations between the three classes ellipticals, spirals and barred spirals, he subsequently introduced evolution and devised his famous tuning fork diagram (Hubble~\cite{h36}). Since then, the most successful galaxy classifications consisted in variations on this scheme (de Vaucouleurs~\cite{devauc}, Sandage~\cite{sandage}, Roberts \& Haynes~\cite{rh}, Kormendy \& Bender~\cite{kb}, van den Bergh~\cite{vdb}) while getting rid of the original evolution significance of this diagram. 

Nowadays, a new revolution is in progress, extragalactic astronomy living a fascinating expansion with the advent of huge and sensitive telescopes. Amazing observed details reveal the complexity and diversity of galaxies, and more precise data on distant galaxies compel us to build the evolutionary history of galaxies as a whole. The nature of the very first objects, still speculative, becomes more and more constrained between initial density fluctuations, observed through the cosmic background at 2.7~K, and the most distant galaxies known, for which distance records are regularly  beaten (Steidel~\cite{steidel}, Pello, Schaerer, Richard, Le Borgne and Kneib~\cite{pello}). Consequently many appellations, supposedly more limited in scope than the Hubble classification, exist. They are also based on appearance using specific observational criteria (radio emission, Lyman break spectral feature, activity in the nucleus, starbursts, dwarf or giant galaxies, etc.).
Traditional classification is thus challenged, both because the number of observed objects rules out the eye-based work still required to determine the morphology of a galaxy, and more fundamentally because this quasi mono-parameter approach is obviously inadequate to encompass the diversity and complexity of galaxies through their evolutions. 

The second approach to classification is global similarity now known as multivariate analysis. It was introduced by
Adanson (\cite{adanson}) in an attempt to correct subjectivity in the choice of classifying characters: Nature should decide which ones are important, not human beings. Multivariate analysis, known as phenetics in biology, thus takes all describing characters available at the epoch of study. It has been an incredibly successful idea widely used in biology until the end of the 20th century and is still very useful in many disciplines. To our knowledge, only a few attempts have been undertaken in astrophysics for a truly multivariate (not considering solely morphology) galaxy classification (see e.g. Whitmore~\cite{whitmore}, Watanabe, Kodaira and Okamura~\cite{watanabe}) while it is widely used for astronomical data analysis (e.g. Feigelson and Babu~\cite{feigelson}, Corbin, Urban, Stobie, Thompson and Schneider~\cite{corbin}).

The third approach is related to the hierarchical organization of biological diversity that was found quite early in the Middle Ages. Later, Linn\'e devised his still used nomenclature accordingly. But this scheme of relationship between species was explained only in the 19th century with the discovery of evolution through natural selection by Darwin (\cite{darwin}). 
Evolution creates diversity and hierarchy is caused by the so-called branching evolution: one species gives birth to descending species (see Fraix-Burnet, Douzery, Choler and Verhamme~2004, hereafter \cite{paperII}). Since a living organism is described by its constituents and their properties, its evolution is nothing else than the evolution of all its characters and their interactions. It is only in the mid-20th century that this fact was used to build a new  methodology to derive phylogenies, called cladistics (Hennig~\cite{hennig}). Essentially all published trees of life now use this methodology (see for instance Stewart~\cite{stewart}, Brower, de Salle and Vogler~\cite{brower}). In this view, two (or more) objects are related if they share a common history, that is they possess properties inherited from a common ancestor. It is important not to be confused between evolution of individuals (genealogy) and evolution of species they represent (phylogeny). Cladistics is only concerned with the latter.

In astrophysics, Hubble remains the only author to have used evolution in the design of his morphological classification scheme. Too often, people try to model and understand evolution \emph{along} the Hubble tuning fork diagram as if this artificial arrangement of the different morphologies is an observational fact. Indeed, astrophysicists roughly understand most of the physical and chemical processes at work in galaxies (including morphology) and are able to model them reasonably well. They now face the problem of synthesising all such knowledge and all such observations.

Because galaxies are defined as independent gravitational groups of stars, gas and dust, their evolution is dictated by the evolution of their constituents. Hence, Fraix-Burnet, Choler and Douzery~(\cite{FCDa}) introduced 'astrocladistics' in an attempt to apply cladistics to galaxies. Van den Bergh (2003, private communication) and Keel(\cite{keel}) also considered this possibility independently. Fraix-Burnet (\cite{fraix}) presents an overview of the progress of astrocladistics which have now tackled several samples of galaxies.

In the present two companion papers (this one and paper II), we present the fundamentals of astrocladistics. This first paper concentrates on the applicability of cladistics to objects like galaxies, emphasizing also the practical course of the analysis. The second paper is devoted to the formalization of the concepts of formation and evolution of galaxies by examining the diversification processes. 

In these two papers, we make use of simulated galaxies from the GALICS project in which galaxy formation and evolution is simulated from a redshift of 30 to 0 (0 being the present time). Even though most of the physics leading from dark matter fluctuations to baryonic objects is still not well understood, such numerical simulations begin to study the entire history of galaxy formation as well as the origin of galaxy diversity. In this first paper, a sample is used to show that chronology can be reconstructed by a cladistic analysis, while with another sample in the second paper it is shown that galaxy diversity can be arranged on a tree-like structure.

In Sect.~\ref{principles} of the present paper, we show that, theoretically, galaxies are suitable objects for a phylogenetic analysis. This is illustrated in Sect.~\ref{cladanal} with the detailed presentation of a cladistic analysis making it obvious that all the ingredients are available for an application to galaxies.
In Sect.~\ref{samples}, after a brief introduction to the GALICS database, we describe how we defined and selected our sample. Results are described in Sect.~\ref{results}, while a discussion can be found in Sect.~\ref{discussion}. Summary and conclusion are given in Sect.~\ref{conclusions}.


\section{Astrocladistics: general principles}
\label{principles}

\subsection{Systematics and cladistics}
\label{systclad}

The two fundamentals of cladistics are in no way related to the kind of objects being studied: (i) diversity generated by evolutionary processes, (ii) hierarchical (tree-like) organization of this diversity due to the branching type of evolution (Wiley, Siegel-Causey, Brooks and Funk~\cite{cladist}, Brower~\cite{brower2000}, \cite{paperII}). On the practical side, cladistics requires objects that can be described by characters for which transformation can be documented. For instance, alongside biology, it is used in linguistics (Wells~\cite{linguistics}), stemmatics (study of ancient books: Robinson \& Robert~\cite{stemmatics}) and even in manufacturing organization (e.g. Tsinopoulos \& McCarthy~\cite{manuf}). 

Innovations appearing in an ancestral species propagate themselves through all its descendants. Cladistics relates objects by identifying these innovations called synapomorphies, i.e. shared derived (=evolved) character states. Reality is more complicated since the same innovations can appear independently at different times in different species and converge through different evolutionary paths (convergences) or even disappear at a later time (reversals). These processes, called homoplasies, bring noise but can be identified in the cladistic analysis.

Cladistics is a methodology, a tool to synthetically visualize informative evolution data as well as hypotheses. It does not reveal the ``true'' evolutionary tree, but merely a possible one given the available data, current knowledge, identified hypotheses and chosen criteria (Wiley et al.~\cite{cladist}, see Sect.~\ref{cladanal}). It has a very powerful interpretive and predictive power regarding the evolutions of all characters because the final evolutionary scenario must be entirely consistent with all the input information. The assumption of branching evolution is thus evaluated in the same process.

Cladistics is not concerned with objects as individuals, but rather with species. It is not an analysis of the genealogy (who is parent of whom) but of the phylogeny (who is the cousin of whom). It is not aimed at identifying the ancestor of a group of objects, because this ancestor is generally not available to the present day observer of living organisms or galaxies. Rather, cladistics groups together objects that share a common ancestor (see e.g. Wiley et al.~\cite{cladist}).

\subsection{On the evolution of galaxies}

Galaxies are independent groups of stars, gas and dust. These constituents, with all their properties (or characters), both \emph{define} and \emph{describe} a galaxy. They evolve by themselves, through interactions with one another and under external perturbations. They \emph{are} the evolution of galaxies and galaxy species (Vilchez, Stasinska and Perez~\cite{evol1}, Sauvage, Stasinska and Schaerer~\cite{evol2}, Hensler, Stasinska, Harfst, Kroupa, and Theis~\cite{evol3}).

The formation of a galaxy is the gathering of these elements in one gravitational entity, each one with its own history. The merging of two galaxies is the mixing together of their constituents whose properties can be seriously modified by the associated perturbations. An interaction of a galaxy with its environment (gas, gravitational potential due to other galaxies or dark matter) often strongly affects its internal constituents (starbursts, accretion or sweeping of gas, collapse onto central black hole, ...): the galaxy might afterwards look significantly different, as it would if it belonged to a new type or class or species. All these formation and diversification processes imply the transmission of the properties of the previous galaxy to the new one, with modification of some of the properties depending on the evolutionary driver. This characterizes a branching evolution as discussed in more detail in \cite{paperII}.

Hence, galaxies are good candidates for a cladistic analysis both because their evolution creates diversity and because the evolution of their fundamental constituents can be observed, interpreted and predicted through observations and physical/chemical models. Such an analysis can tell us the evolutionary relationships among different classes of galaxies, through a synthesis of the evolutions of \emph{all} the physical and chemical processes as contained in the input characters. These are observable descriptors of the fundamental constituents. In astrophysics, the characters are to be found in spectra: absolute luminosities, colours or flux ratios, kinematic information, spectral line properties. Morphology is not adequate because it is subjective (essentially determined by eye) and qualitative. More importantly, it is included in kinematical characteristics because the apparent morphology of a galaxy essentially reflects the 3-dimensional motions of the stellar population but is always observed as a 2D projection only. 

What we call ``classes'' or ``species'' will be defined \emph{after} the analysis of a large sample of galaxies is done, that is after the evolutionary relationships are better understood. 
In cladistics, two taxa are grouped together if they share some common derived character states (see Sect.~\ref{matrcod}) which are supposed to have been inherited from a common ancestor. Astrocladistics will thus lead us toward a new taxonomy, that is toward a new classification and nomenclature. Before that can be made, and particularly in this paper, each individual galaxy is considered as a representative of a class (see Sect.~\ref{samplechoice}). We also use the word ``class'' in a generic manner, and avoid the word ``type'' because it is inevitably linked to the Hubble morphological classification.


\section{Phylogenetic analysis of galaxies using cladistics}
\label{cladanal}

This section is intended to outline the basic steps of a cladistic analysis, with emphasis on its general application to galaxies, and giving specific parameters used in the analysis of the sample introduced in Sect.~\ref{samples}. It is not a thorough presentation of this methodology which can be found in many places (e.g. Wiley et al.~\cite{cladist}).

\subsection{Outline}
\label{outline}

The principal qualities of a cladistic analysis are objectivity and transparency.
Practically, the objects under study are described by evolutionary characters (Sect.~\ref{matrbuild}) for which at least two states are defined (Sect.~\ref{matrcod}): one is said to be ancestral, the other one is said to be derived. Most often, this evolution orientation is known for a very few characters only, increasing the number of possible phylogenies. The derived state corresponds to an innovation in the evolution and is assumed to have been acquired by an unidentified ancestor.
In cladistics, objects are grouped from the \emph{derived} character states they share, encompassing the ancestor and all its descendants after transformation from the ancestral state. The basic ingredient to the analysis is thus a matrix scoring for all taxa the states for all characters (Sect.~\ref{matrcod}). 

\begin{figure}
   \centering
   \includegraphics[width=9 true cm]{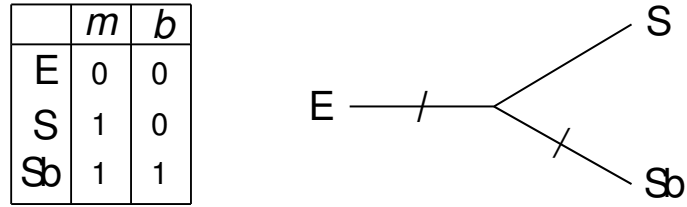} 
      \caption{Illustration of how the Hubble tuning fork diagram can be obtained from a cladistic analysis. \textit{m} and \textit{b} are the two characters ``morphology'' and ``bar'' respectively. ``0'' stands for absence (of spiral arm or bar), ``1'' meaning presence. The slashes on branches indicate a change in one character state. For instance, to go from type E to type S, one needs to modify the state of character \textit{m} (from 0 to 1). This tree is said to have two steps (two character state changes altogether).
              }
         \label{hubbleclad}
   \end{figure}

Because cladistics assumes branching evolution, the evolutionary relationships are represented on a tree or cladogram. 
The process for finding the trees is very basic and can be done by hand. This is illustrated on Fig.~\ref{hubbleclad} which schematically demonstrates how the Hubble diagram can be obtained from a cladistic analysis. Let us consider three galaxy classes and two characters with two states: morphology (spiral or elliptical) and bar (present or absent). Assuming these states represent evolution and considering the changes of character states necessary to evolve from one type to the other, we can build an unrooted cladogram which happens to be exactly the Hubble tuning fork diagram. To root the cladogram, one should decide, at least for one character, which state is ancestral (Sect.~\ref{outgroup}). For instance, Hubble thought that ellipticals become spirals with time. Hence he would have defined '0' as ancestral state for character $m$. Nowadays, the reverse is preferred, which would make '0' the derived state.

The goal of astrocladistics can be viewed as to extend this exercise to all possible descriptors (for better objectivity) and many more classes of galaxies (for better coverage of galaxy diversity). This is presented in Sect.~\ref{besttree}. 

Finally, there are some statistical methods to assess the robustness of the result trees (Sect.~\ref{assess}) which must be used before trying a thorough interpretation (Sect.~\ref{interpr}).

\subsection{Choice of a sample}
\label{samplechoice}

The aim of astrocladistics is to group objects from homologies. But there are millions of galaxies in the Universe, the nature and composition of the very first objects are unknown (groups of stars, gas overdensities ?), so that there might well be several different kinds of initial seeds of galaxies. Thus picking up galaxies that might have relatively close evolutionary relationships is somewhat like a fishing experience at such an exploratory stage.

There are at least two possible approaches: i) use probability estimates of close relationships within the original sample and then group objects accordingly before performing the cladistic analysis (e.g. Wiens~\cite{wiens} and references therein), ii) assume one object - one class and let the cladistic analysis identify groups from evolutionary considerations. For these first developments of astrocladistics, we have favoured the second approach to avoid possible biases of pre-classification introduced by the first one. Later on, it would be certainly interesting to compare the two. We also try to begin with galaxies that belong spatially to the same group or cluster because they more probably have derived from the same kind of initial seeds and evolved in the same global environment. This is however not the case in the two present companion papers that use simulated galaxies.

Even if millions of galaxies are currently known, a tiny fraction of them are described with more than a dozen observables. Currently,  between about 50 to 100 characters can be obtained for a few hundred galaxies which are all quite close to us, hence very evolved. Thanks to space and big aperture ground-based telescopes, this is changing rapidly. A future development will be to extract from the medium or high resolution spectra the adequate information to be used by  methodologies such as astrocladistics. Nevertheless, since galaxies might not be as complex objects as living organisms, it is thus highly probable that the numbers of pertinent characters will not be significantly more than one or two hundred. This is still much higher than in Hubble's time, and large enough to define many classes.
As technology improves, as knowledge increases, new characters become available requiring analyses of the same groups of objects to be redone periodically. Hence, classifications must evolve and are never definitive. 

\subsection{Building the matrix}
\label{matrbuild}

In order to remain as objective as possible, it is preferable not to choose characters a priori, but rather to take all the ones available at the time of study and let the analysis reveal possible inconsistencies in character behaviours. Ideally, characters should be independent regarding evolution. One first obvious reason is to avoid redundancy of information and overweight of a single evolutionary process. A second less obvious reason is to ensure a better statistical significance of the resulting evolutionary scheme by minimizing possible conflicts between character evolutions. Also, the use of the parsimony criterion (as will be described in Sect.~\ref{besttree}) emphasizes characters that do not change too much nor too erratically. However the requirement of character independence is not so easy to satisfy \textit{a priori} and should be re-analysed afterwards. Note that two characters can be independent and \emph{apparently} correlated (for instance metallicity and mass of galaxies can both increase with evolution but are not necessarily physically related). Differential weights can be given to characters, but this will not be considered in this series of papers.

Two unique particularities of astrophysical observations are not usual to cladists: uncertainties and upper/lower limits. Because they are quite informative and common, they should be included. Physical data are meaningless if they are not accompanied by an estimation of the accuracy of the measurement. These can be treated in the analysis either by a weight matrix or in the coding process (see following section). Upper and lower limits generally result from limited sensitivity of the detectors and/or conditions of observations (bad seeing, background sky brightness, source confusion, etc.). Hence, in a given sample and for a given character, they do not always correspond to extreme ends of ranges of values. They also do not reveal the distribution of values below (above) the lower (upper) limit. Thus, when possible, they can be in one bin at either end of the codings. 

\subsection{Coding the matrix}
\label{matrcod}

Astrophysical observables are in general continuous values. There are strong debates on whether or not such data should be used in biology (e.g. Rae~\cite{rae}), but astronomers have no choice. Yet, we think their use is totally legitimate in the case of galaxies because the change of characters is mainly gradual and totally reflects evolution. That being said, there remain several possibilities to code such values. Thorough investigations will have to be performed in the future.

For a cladistic analysis, at least two evolutionary states should be identified for each character: an ancestral state and a derived state. Depending on the character, depending on the sample, characters can be coded into several states. Continuous data should be binned, and unknown values are allowed.

Upper or lower limits are not explicitly treated as such in current cladistic software. Nevertheless they can be considered as real values if they correspond to upper or lower bins and possibly be attributed a lower weight. Uncertainties can be treated in two ways. First, bins can be chosen to be significantly larger than the error bars of the data. Second, it is always possible to do several analyses with slightly different codings and compare results. No upper/lower limits nor error bars are present in our GALICS sample, so we do not discuss this point any further.

For the simulated sample of this paper, we attributed eight evolutionary states to each character by regularly binning the corresponding range of values among all galaxies. For photometric characters, three kinds of such coding have been performed : using flux values, magnitudes, and colours with respect to the Johnson K band, and comparing the results (Sect.~\ref{fmc}).

\subsection{Introducing additional constraints} 

The more information is given initially, the more constrained and robust the result will be. For a given group of taxa, the more characters the better. But any knowledge on the evolutions of these characters is also extremely useful to help eliminate some evidently impossible evolutionary scenarios. This also constitutes a means to test different hypotheses or models. 

In this paper, we impose that all characters evolve smoothly with time, that is they are all supposed to be ordered: changes between two adjacent states are more parsimonious (see Sect.~\ref{besttree}) than between distant ones. In other words, big jumps are supposedly less probable than gentle character evolution. This hypothesis is physically sensible in the case of galaxies and we found that it significantly improves the robustness of trees.

\subsection{Defining the outgroup}
\label{outgroup}

Defining evolutionary states only yields an unrooted tree because the arrow of time is not indicated. Time behaviour of at least some characters is required to root the cladogram. This can be introduced in the additional constraints and/or by defining a comparison group ('outgroup') which is a real or hypothesized taxon having some identified ancestral states. This outgroup should be outside the studied sample and share a common ancestor with it, even if in practice it  is often regarded as part of the sample. Hence it should be neither too far nor too close in the evolutionary diversification. The choice of the outgroup is always delicate and rarely unique. The detailed interpretation of the resulting phylogeny for the group of study depends on this choice, but the classes of possible phylogenies are revealed even on an unrooted tree.

In astrophysics, determining potential outgroups for the sample under study is at present nearly impossible. In general, it could be possible to build an artificial outgroup because some characters have a known global evolution through the lifetime of the Universe. For instance, the metallicities and the masses of galaxies are expected to increase because stars gradually transform light atomic elements (hydrogen, helium) into heavier ones (oxygen, carbon) and gravity makes galaxies bigger with time via accretions and mergers. But locally, an accretion of a big cloud of hydrogen gas can diminish the average metallicity of a galaxy, and through interactions galaxies can be torn apart into smaller pieces.

In this paper, we decided not to root the trees, because this part of our work is intended to demonstrate that physical descriptors of galaxies do trace their evolution and show how astrocladistics reconstruct the chronology. For display purposes only, we assume that epoch 1 galaxies are closer to the ancestor common to the entire group. This is equivalent to defining an outgroup, but here it is an arbitrary choice which we consider sufficient for the present work.

\subsection{Finding the best trees}
\label{besttree}

For a given matrix (a set of taxa and coded characters), the number of possible trees is huge and grows as $(2n-3)!/2^{n-2}(n-2)!$ where $n$ is the number of objects (Swofford, Olsen, Waddell and Hillis~\cite{swofford}).
For more than 4 objects with 4 characters, the process of finding trees is not tractable by hand any more (see Sect.~\ref{outline}) and computers are required. There are several software packages available. We used the PAUP4b10* package (Swofford~\cite{paup}) on Linux PC computers to perform all calculations shown in this paper.

To choose among the huge amount of possible trees, a usual criterion is maximum parsimony: the total number of character state changes (so-called 'steps') on the tree is minimized. This also minimizes the number of homoplasies (convergences or reversals in character evolution) which perturbs the grouping in evolutionary classes. If several most parsimonious trees are found, then a 'consensus' tree can be built.

In this paper, because the sample was arbitrarily chosen, galaxies have a chance to be too distant in the evolutionary diversification. It is also possible that there exist different evolutionary pathways, with several very different ancestors. This prevents a good relationship to be found and leads to a largely unresolved tree. Our strategy was to find a sub-sample that yields a fully resolved tree. For this purpose, we eliminated taxa one by one, running the analyses and comparing the number of bootstrap values above 50\% (Sect.~\ref{assess} for details on bootstrap method) on all the resulting trees: the best resolved tree was kept and the corresponding taxon definitively eliminated before the same process started again with the new sub-sample. We thus ended up with a robust tree concerning objects that can be said to be evolutionarily related.

In these two companion papers, we further attempted an analysis with all the eliminated galaxies since the number of characters should be higher than the number of taxa to obtain a good resolution of the phylogeny. If a robust tree is found, this would mean that it might either be a kind of redundant object with respect to the first tree or constitute another monophyletic group with a different ancestor. Another strategy, not considered in these two companion papers, would be to place the eliminated taxa onto the first tree. This could imply defining groups and would allow us to extend the cladistic analysis to other objects to build progressively a phylogenetic classification of larger samples of galaxies.

\subsection{Assessing the phylogenetic signal in the data}
\label{assess}

After the most parsimonious trees are found, some statistical tests are performed in order to assess the robustness of the results. Ideally, the phylogenetic signal (evolutionary information) should not depend on too few dominant characters because of possible errors or uncertainties in the data, misbehaviours of these characters or too high a dependence on current level of knowledge and observational techniques. The bootstrap method (Felsenstein~\cite{Felsenstein}, Efron \& Tibshirani~\cite{efron}) analyses the robustness of the tree to noise in the sample through the percentage of times a node is found among results from resampled matrices (1000 in this paper). These matrices are identical in size to the original one, but each character column is randomly selected from the entire initial set. Hence, a resampled matrix can have several times the same character and lack some of them. Bootstrap is therefore a way to put random weights on individual characters, and then evaluate the ability of the tree-building method to recover the initial nodes despite these differential character weightings.
The Decay (or Bremer) index indicates how many more steps on the tree are needed to destroy the corresponding branch (Bremer~\cite{bremer}). It measures the robustness of the supposedly-best tree which has been selected using the parsimony criterion. 

Two other indicators are generally given with each tree and can be computed for each character or globally. The Consistency Index (CI) measures the difficulty of fitting a data set on a particular tree: it is the ratio between the minimum number of possible steps (given by the data matrix) and the total number of steps of the tree. It is always lower than 1 and a perfect phylogeny, i.e. without homoplasy in the characters, analysis would give a CI of 1. This indicator can thus be used to examine the behaviour of a given character on a particular tree. The Retention Index (RI) measures the level of similarities in the tree. In a sense, it measures for the same matrix, the distance of the result tree to the worst case (totally unresolved tree due to total lack of phylogenetic signal) and to the best one (perfect phylogeny with the result tree). RI behaves like CI, being larger in good situations (Wiley et al.~\cite{cladist}).

\subsection{Interpreting the cladograms}
\label{interpr}

Two branches on the tree are linked by a node which represents the collection of shared derived characters due to common ancestry. A long branch with several nodes is called a lineage. All descendants characterized by unique derived characters inherited from their common ancestor constitute a monophyletic (phylogenetic) group or clade. This is the basis of a phylogenetic inference based on a cladogram.

Interpretation of the result is done by projecting the coded characters onto the branches of the cladogram. 
Self-consistencies and estimation of global consistency with input data, hypotheses and other current thoughts on galaxy evolution and their physics, helps one to conclude on the validity of the evolutionary scenario proposed by the cladistic analysis. Groupings find historical explanations in the projected character evolutions along branches, and anomalies can be identified. Finally predictions may come out for objects on the tree that have missing data.


\section{A test sample: simulated galaxies without mergers}
\label{samples}

In this first paper, to simplify as much as possible the cladistic analysis and the interpretation of the results, a sample of galaxies that have never had any merger events was selected. Our goal here is to demonstrate that from observables describing galaxies, that is characterizing their basic constituents, it is possible to reconstruct the historical relationships among a sample of galaxies. \cite{paperII} describes in detail the processes of formation and diversification of galaxies and enlightens our choice some more.

\subsection{Brief presentation of GALICS}

GALICS (Galaxies In Cosmological Simulations) is a hybrid model for hierarchical
galaxy formation studies, combining the outputs of large cosmological N-body simulations
with simple, semi-analytic recipes to describe the fate of the baryons within dark matter haloes (Hatton, Devriendt, Ninin, Bouchet and Guiderdoni~\cite{galics1}). A galaxy appears when the density of baryonic matter is above a given level corresponding to the resolution of the simulation. As hot gas cools and falls to the centre of these haloes, it settles
in a rotationally supported disc. Galaxies remain disc structures unless mergers or instabilities occur, in which case simple recipes are used to develop a bulge and a burst components. Hence a galaxy is described by these three components each one having its own parameters of geometry, dynamics, masses (stars and gas), metallicity and photometry from the ultraviolet to the far infrared.
Stellar and chemical evolution is modelled, but no interaction between galaxies are considered.

Each galaxy is identified by a specific number at each timestep of the simulation. Each galaxy is the product of one or more galaxies of the previous step and one or more evolutionary processes that occurred since the previous step. The entire genealogy of each galaxy is thus known. It is then possible to select galaxies that have never been the product of the merging of two or more galaxies.

\subsection{Problematics of selecting a sample}
\label{diffchoi}
It must be mentioned that choosing galaxies for a phylogenetic analysis is not a trivial problem at this preliminary stage. Taking objects too close in the evolution process hampers the diversity resolution. Conversely, if galaxies are too different (i.e. too much diversified), it is difficult to find the evolutionary relationship. In both cases, it is difficult to find a robust tree. The problem lies in the definition of similarity which is here taken in the evolutionary sense and can be really understood only \emph{after} a cladistics analysis is made. Comparing objects in the traditional way (with a few apparent characters) or with a multivariate distance analysis is not adequate because of probable homoplasies (see Sect.~\ref{systclad}). One might hope to find galaxies identically described based on coded characters. Unfortunately, a few tens of characters having a few states already makes a substantial number of possibilities so we have not yet encountered such cases.

The problem would be certainly much simpler with simulated galaxies. However, we wanted to train ourselves by considering the simulation as if it was the real Universe, with no genealogical information available to the observers. This information was used only to select non-merger galaxies in the database and compare with our results afterwards. 

As described in Sect.~\ref{besttree}, a huge number of possible trees exist for a given set of objects and characters. The analysis is thus very CPU time consuming and cannot be reasonably done with too many objects. We consider that about 50 is a good compromise with our present knowledge of cladistic galaxy classification, and with the number of currently available descriptors (characters). As already mentioned in Sect.~\ref{samplechoice} and as discussed in \cite{paperII}, at this stage of astrocladistics, each galaxy represents a class to be defined later on.

\subsection{The sample of galaxies without merger}

We chose 10 galaxies at 5 epochs (simulation step 30, 40, 50, 60 and 70 corresponding to redshift of 3, 1.9, 1, 0.4 and 0). These galaxies were arbitrarily chosen among all galaxies born at redshift 3 (simulation step of 30) and having experienced no merging during the time spanned by the 5 epochs. They are representative of classes, and named from An to Jn where n is the epoch (1 to 5), 1 being the most ancient one and 5 the present (redshift = 0).

In the GALICS database 91 characters are available to describe these galaxies and listed in Table~\ref{tabchar}. Two characters are global (bolometric luminosity and infrared luminosity), the other ones describing the three components of GALICS galaxies: the disc (31 characters), the bulge (27) and the burst (31). Most ($3 \times 23$) are broad band magnitudes ranging from U to 500~micron. As explained in Sect.~\ref{fmc}, all magnitudes are relative to the K band, this last value giving the relative heights of the spectra or relative brightnesses between galaxies.
The dynamical time $tdyn$ is the time taken for material at the half-mass radius to reach the opposite side of the galaxy (disk component) or its centre (bulge and burst components), whereas the star formation rate is derived from this dynamical time, the mass of the cold gas and a prescribed star formation efficiency, either instantaneous at the last time substep of the simulation or averaged over the last step (Hatton et al.~\cite{galics1}).

\begin{table}
	\begin{center}
		\begin{tabular}{rlrlrl}
 1& bol\_lum              & 2 &IR\_bol                 &   &                        \\
 3& disc\_mcold           &   &                        & 15& burst\_mcold           \\
 4& disc\_mstar           & 11& bulge\_mstar           & 16& burst\_mstar           \\
 5& disc\_mcoldz          &   &                        & 17& burst\_mcoldz          \\
 6& disc\_rgal            & 12& bulge\_rgal            & 18& burst\_rgal            \\
 7& disc\_speed           & 13& bulge\_speed           & 19& burst\_speed           \\
 8& disc\_tdyn            & 14& bulge\_tdyn            & 20& burst\_tdyn            \\
 9& disc\_av\_sfr         &   &                        & 21& burst\_av\_sfr         \\
10& disc\_inst\_sfr       &   &                        & 22& burst\_inst\_sfr       \\
  &                       &   &                        &   &                        \\
23& disc\_JOHNSON\_U*     & 46& bulge\_JOHNSON\_U*     & 69& burst\_JOHNSON\_U*     \\
24& disc\_JOHNSON\_B*     & 47& bulge\_JOHNSON\_B*     & 70& burst\_JOHNSON\_B*     \\
25& disc\_JOHNSON\_V*     & 48& bulge\_JOHNSON\_V*     & 71& burst\_JOHNSON\_V*     \\
26& disc\_JOHNSON\_R*     & 49& bulge\_JOHNSON\_R*     & 72& burst\_JOHNSON\_R*     \\
27& disc\_JOHNSON\_I*     & 50& bulge\_JOHNSON\_I*     & 73& burst\_JOHNSON\_I*     \\
28& disc\_JOHNSON\_J*     & 51& bulge\_JOHNSON\_J*     & 74& burst\_JOHNSON\_J*     \\
29& disc\_JOHNSON\_K      & 52& bulge\_JOHNSON\_K      & 75& burst\_JOHNSON\_K      \\
30& disc\_SCUBA\_850mic*  & 53& bulge\_SCUBA\_850mic*  & 76& burst\_SCUBA\_850mic*  \\
31& disc\_UV\_1600Ang*    & 54& bulge\_UV\_1600Ang*    & 77& burst\_UV\_1600Ang*    \\
32& disc\_UV\_1500Ang*    & 55& bulge\_UV\_1500Ang*    & 78& burst\_UV\_1500Ang*    \\
33& disc\_IRAS\_100mic*   & 56& bulge\_IRAS\_100mic*   & 79& burst\_IRAS\_100mic*   \\
34& disc\_IRAS\_12mic*    & 57& bulge\_IRAS\_12mic*    & 80& burst\_IRAS\_12mic*    \\
35& disc\_IRAS\_25mic*    & 58& bulge\_IRAS\_25mic*    & 81& burst\_IRAS\_25mic*    \\
36& disc\_IRAS\_60mic*    & 59& bulge\_IRAS\_60mic*    & 82& burst\_IRAS\_60mic*    \\
37& disc\_ISOCAM\_15mic*  & 60& bulge\_ISOCAM\_15mic*  & 83& burst\_ISOCAM\_15mic*  \\
38& disc\_PACS\_110mic*   & 61& bulge\_PACS\_110mic*   & 84& burst\_PACS\_110mic*   \\
39& disc\_PACS\_170mic*   & 62& bulge\_PACS\_170mic*   & 85& burst\_PACS\_170mic*   \\
40& disc\_PACS\_75mic*    & 63& bulge\_PACS\_75mic*    & 86& burst\_PACS\_75mic*    \\
41& disc\_SIRTF\_3\_6mic* & 64& bulge\_SIRTF\_3\_6mic* & 87& burst\_SIRTF\_3\_6mic* \\
42& disc\_SIRTF\_8\_0mic* & 65& bulge\_SIRTF\_8\_0mic* & 88& burst\_SIRTF\_8\_0mic* \\
43& disc\_SPIRE\_250mic*  & 66& bulge\_SPIRE\_250mic*  & 89& burst\_SPIRE\_250mic*  \\
44& disc\_SPIRE\_350mic*  & 67& bulge\_SPIRE\_350mic*  & 90& burst\_SPIRE\_350mic*  \\
45& disc\_SPIRE\_500mic*  & 68& bulge\_SPIRE\_500mic*  & 91& burst\_SPIRE\_500mic*  \\
		\end{tabular}
	\end{center}
	\caption{List of characters. \textit{mstar}, \textit{mcold} and \textit{mcoldz} stand respectively for the masses of stars, gas and metals. \textit{rgal} is the component radius, \textit{speed} its rotation speed, \textit{tdyn} the dynamical time, \textit{av\_sfr} and \textit{inst\_sfr} respectively average and instantaneous star formation rates. See text for more details. Magnitudes (characters 23 to 91), for different broad band filters at different wavelengths,  that are starred, are relative to the K band of each component.}
	\label{tabchar}
\end{table}


\section{Results}
\label{results}

\subsection{Flux, magnitudes or colours?}
\label{fmc}

Even if all characters should be used in a cladistics analysis, it is important to avoid obvious biases. In astrophysics, an invaluable source of information lies in the spectra, or here in the broad band magnitudes. But since a bright galaxy is very probably bright at nearly all wavelengths, using these characters crudely would give too much weight to the galaxy luminosity. Rather, colours or relative luminosities are much more informative about the different components in a galaxy and for evolution. We found that they provide better robustness to the final tree, which tends to validate this statement. We will only consider the colour based result in the rest of this paper.

Magnitudes are merely transcription (logarithmic) of fluxes. The choice among the two implies somewhat different distributions of objects into character codes and allows for different resolution of diversity. We believe that there is no general rule on this point and both should be tried. On the present sample, the results were not very different in terms of relationships, and because trees are found to be slightly more robust, we preferred coding logarithmic values (magnitudes).

\subsection{Reconstructing the correct chronology}
\label{corrchro}

\begin{figure}
   \centering
   \includegraphics[width=10 true cm]{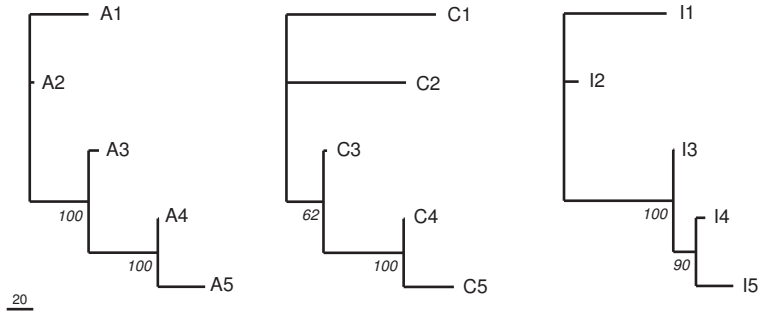} 
      \caption{Evolution of the individual galaxies (A to J) obtained by 10 independent analyses of matrices with 5 objects corresponding to the 5 epochs. All cladograms are most parsimonious trees. The cladograms that are not shown are all similar to the one for A. Numbers are bootstrap values. Branch lengths are here proportional to the number of character state changes (tick mark indicates 20 steps) and are not directly related to a timescale.
              }
         \label{indivtree}
   \end{figure}

Is a character evolution based analysis such as cladistics able to figure out the correct chronology of galactic evolution? It is possible to give the answer thanks to the simulated galaxies of GALICS chosen in this work. We considered each of the 10 galaxies A to J at the 5 epochs, building 10 matrices with 5 objects. Analyses were then performed independently for each matrix. In these conditions, the cladistic analysis finds excellent phylogenies as illustrated in Fig.~\ref{indivtree} for galaxies A, C and I. All trees do have the correct chronology, and all have bootstrap values of 100 except for galaxies C and I which have slightly lower indexes. 

The cladograms show that galaxies at epoch 5 are the most diversified when compared to galaxies at epoch 1. Branch length represents number of character state changes, and are not directly related to a time scale. It should be noted that the results are obtained with all characters. As will be seen in Sect.~\ref{whole}, burst characters introduce noise and are certainly not very relevant for galaxy evolution. Still, the results on the 10 sub-samples are excellent and could only be better yet without burst components. Here the capability of astrocladistics to find the right history is thus demonstrated on a concrete basis.

\subsection{History of the whole sample keeping all characters}
\label{whole}

   \begin{figure}
   \centering
   \includegraphics[width=18 true cm]{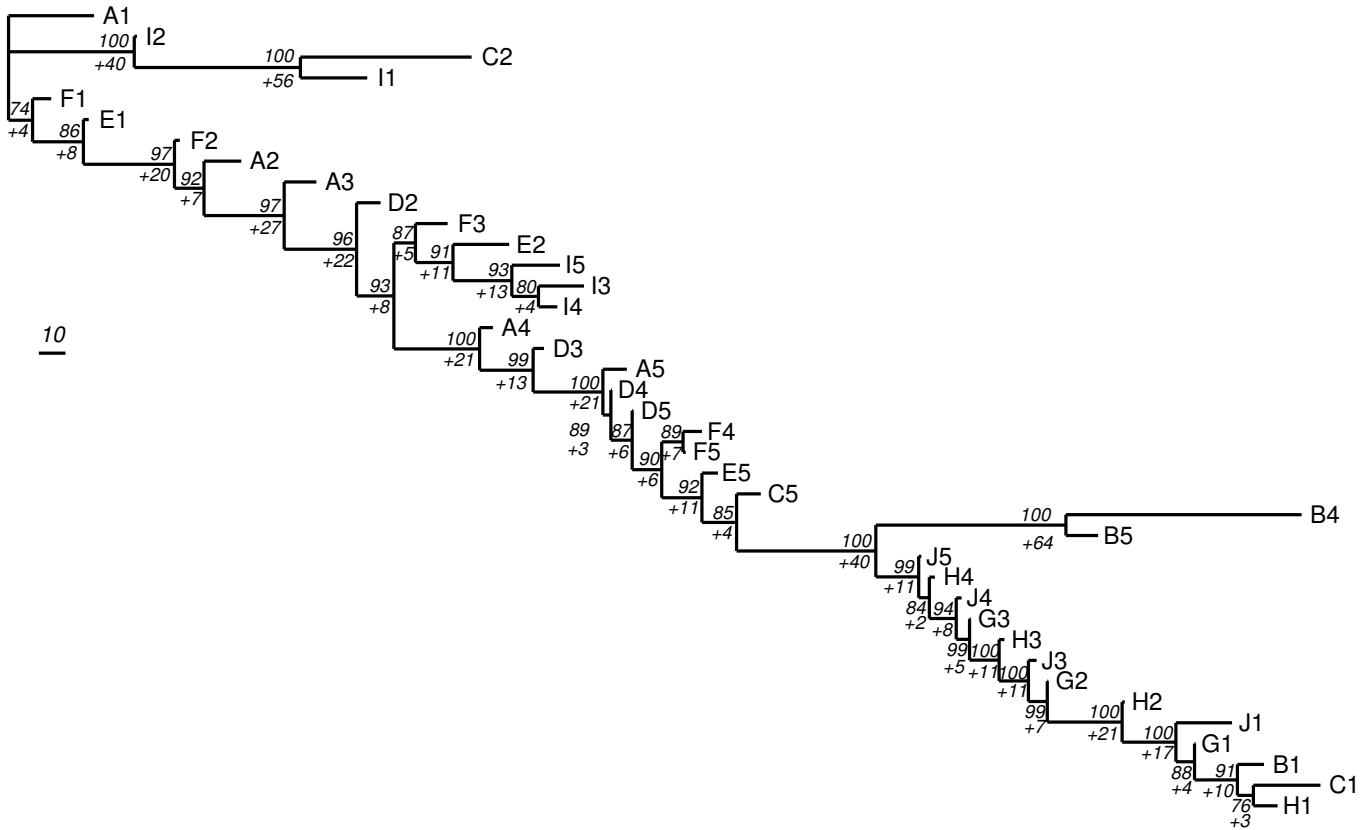} 
      \caption{Most parsimonious tree obtained with the whole sample and all characters after removal of 11 galaxies: B2, B3, C3, C4, D1, E3, E4, G4, G5, H5 and J2. Numbers to the left of each node are bootstrap (above) and decay (below with a plus sign) values. The number of steps (character state changes) for this tree is 1167 (scale bar on the left of figure indicates 10 steps), CI=0.49 and RI=0.84 (see Sect.~\ref{assess}).
              }
         \label{firstclad}
   \end{figure}

In the real world, we do not have the information on common ancestry. Hence, we have to hypothesize it and proceed with the entire analysis in order to test this assumption by trying to find the best tree. As described in Sect.~\ref{besttree}, our procedure was to exclude objects one by one until we find optimal bootstrap values on the result tree. Finally, 11 galaxies were so removed, and we obtained the cladogram shown in Fig.~\ref{firstclad} with the 39 remaining galaxies. Bootstrap values and decay indexes are excellent. 

   \begin{figure}
   \centering
   \includegraphics[width=18 true cm]{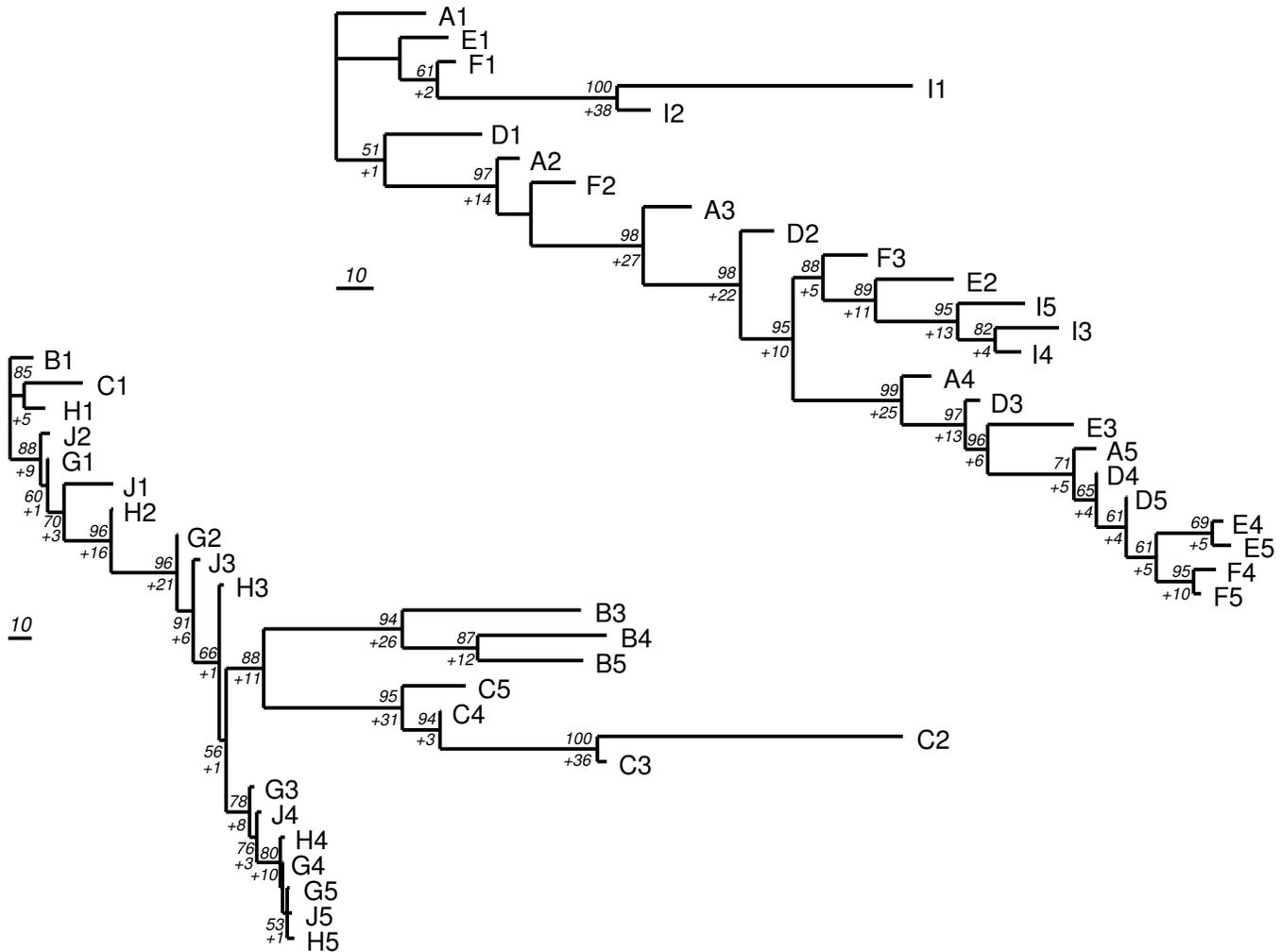} 
      \caption{Most parsimonious trees obtained considering the whole sample contains two groups (ADEFI and BCGHJ) with two different ancestors. Only B2 has been removed to obtain the cladogram on the right. Numbers to the left of each node are bootstrap (above) and decay (below with a plus sign) values. Scale bar on the left of each cladogram indicates 10 steps. Tree for ADEFI: steps=701, CI=0.50, RI=0.80. Tree for BCGHJ: steps=791, CI=0.69, RI=0.81 (see Sect.~\ref{assess}).
              }
         \label{twoanc}
   \end{figure}
   
Going downward along the cladogram, it is possible to see that galaxies order themselves correctly in chronology from A1 taken as the root, to J5 after which the trend goes the other way to end up with C1 and H1 as the most diversified with respect to A1. Changing the root to C5 or B4-B5 would make a tree diverging into two lineages. Both would be chronologically reversed (from epoch 5 to epoch 1). Another option is that there could be two ancestors, A1 and C1 for instance, making the two lineages to merge around C5-B4. The picture would not be a tree anymore and in this case, separate analyses of the two lineages should be done (see Sect.~\ref{twogroups}).

To understand Fig.~\ref{firstclad} and make a reasonable choice for the root, it is useful to look at character evolution, and momentarily assume that the number of character state changes (represented by branch lengths) is grossly proportional to time. As can be seen in Table~\ref{charcod} (last 23 columns on the right), it is noticeable that galaxies B, C, G, H and J have no burst component (code '7' in the table) at epoch 1 and all are at the bottom of the tree. Note also that G, H and J have no burst component at all epochs. The cladogram seems to indicate that the burst luminosity should rather smoothly decrease with time. Indeed, bursts are by definition temporary events provoked by instabilities in the disk or accretion of gas. They can thus appear and disappear at nearly any time as is the case for galaxy B2 and C2. 

\subsection{Two groups and two different ancestors?}
\label{twogroups}

At this stage of the analysis, two options are possible. The first one is to assume that, because galaxies ADEFI and BCGHJ are born with different burst components, they could have two different ancestors. The same analyses have been performed on each one of these subgroups. The resulting cladograms are shown on Fig.~\ref{twoanc}. They are both well resolved, especially because no galaxy removal for optimization was necessary except only for B2. There are one or two little supported nodes on each tree (bootstrap less than 60), but the other nodes show that the two ancestor hypothesis is a very plausible interpretation. By the way, the chronology is perfectly respected in both diagrams.

The second option is to remove burst characters and is detailed in the next section.

\subsection{Removing burst characters}
\label{woburst}

   \begin{figure}
   \centering
   \includegraphics[width=15 true cm]{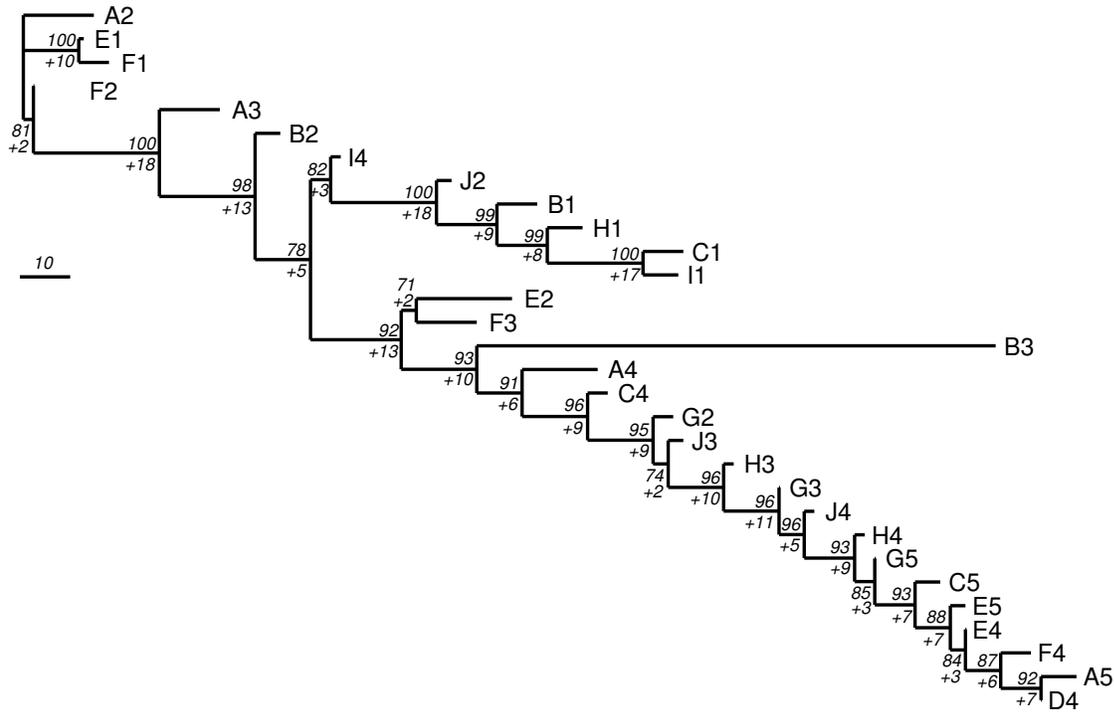} 
      \caption{Most parsimonious tree obtained without burst characters and through optimizing the result by removing one galaxy after another. Numbers to the left of each node are bootstrap (above) and decay (below with a plus sign) values. Steps=542 (scale bar on the left of figure indicates 10 steps), CI=0.50 and RI=0.85.
              }
         \label{bestclad}
   \end{figure}

   \begin{figure}
   \centering
   \includegraphics[width=15 true cm]{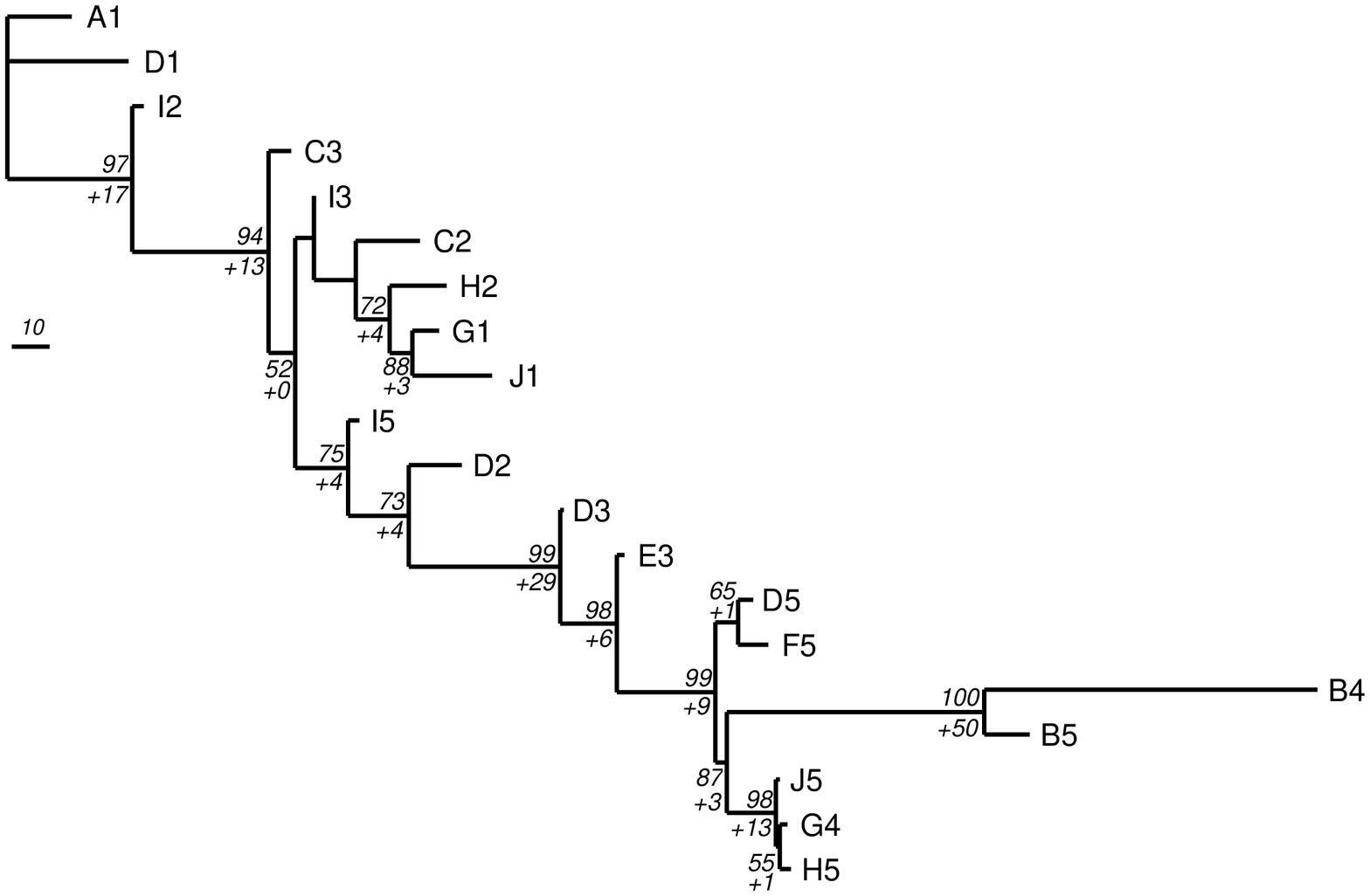} 
      \caption{One of the two most parsimonious trees, differing only in the position of C2, for the excluded galaxies of Fig.~\ref{bestclad} obtained without burst characters. Numbers to the left of each node are bootstrap (above) and decay (below with a plus sign) values. Only nodes with bootstrap higher than 50\% are indicated. Steps=565 (scale bar on the left of figure indicates 10 steps), CI=0.66, RI=0.80. 
              }
         \label{exclad}
   \end{figure}

Since burst characters are doubtful indicators of galaxy evolution, we have done the analysis of the entire sample (50 galaxies) without them (characters 15-22 and 69-91 of Table~\ref{tabchar} were ignored). Among the 60 remaining characters, the bulge photometry ones (46-68) are identical in all galaxies but three (B3, B4, B5, see Table~\ref{charcod}). This makes the total number of significantly discriminant characters somewhat too low to hope to obtain a very robust cladogram for the 50 galaxies.

By the same optimization procedure described in Sect.~\ref{besttree} and used in Sect.~\ref{whole}, 20 galaxies were removed, leaving an excellent cladogram with 30 objects (Fig.~\ref{bestclad}). The 20 excluded galaxies were analysed as well, and the cladogram in Fig.~\ref{exclad} has been obtained without any optimization. The result is excellent, bootstraps are high and the true chronology well respected. There is only one weakly supported node (bootstrap of 52) indicating an unresolved relationship or a few objects too diversified to fit in this group. 

We have thus identified two groups of galaxies, whose cladograms are globally better supported than those obtained from the two ancestor hypotheses of the previous section. Here also we have two groups, but they have not been chosen a priori. As said above, the rather low ratio of number of characters versus number of taxa prevents us from drawing conclusions on common ancestry of these two groups. To definitively answer the question, one would need more discriminant characters, or begin defining classes by analysing behaviour of all characters on the cladograms of Fig.~\ref{bestclad} and Fig.~\ref{exclad}. This is beyond the scope of this paper.

\section{Discussion}
\label{discussion}

Among the different results presented in this paper, those shown in Fig.~\ref{bestclad} and Fig.~\ref{exclad} (Sect.~\ref{woburst}) are clearly the most satisfactory because they are less affected by a priori subjective choice and the evolutionary scenario represented on the cladograms is astrophysically plausible. On the contrary, the analysis using all characters (Sect.~\ref{whole}) is plagued by doubt on burst characters as galaxy evolution indicators. The other results (Sect.~\ref{corrchro} and Sect.~\ref{twogroups}) heavily depend on our a priori knowledge of lineages available thanks to the simulations. They thus seem very artificial and cannot be representative of a real data set.

This illustrates the principal strength of astrocladistics in classifying galaxy diversity. The resulting cladogram is objectively obtained, and can be accurately discussed on this basis. Particular points of debate can be:
\begin{enumerate}
	\item character coding: several choices are possible, different results can be compared. Character values are all quantitative, influence of measure uncertainties can be examined.
	\item evolution of characters: some knowledge or hypotheses can be put here, principally intended to increase tree robustness.
	\item character weighting: extreme care should be taken when imposing weights on characters, probably useful when only upper/lower limits are known, or when uncertainties are large.
	\item choice of outgroup: quite a difficult task because it is generally difficult to find the right one, especially at this stage of astrocladistics. 
\end{enumerate}

The main result of the present paper is that the true chronology is easily found. This proves that (simulated) observables used here to describe galaxies are certainly representative of their evolution, that is of the evolution of their fundamental constituents. This is a clear demonstration that astrocladistics is well-founded. It is essential to note that these observables are available in real galaxy catalogues.

Once again, galaxies here represent classes, or species, and this should be kept in mind when reading the cladograms. They are not considered as individuals, we are not visualizing the development of an object through time, but rather the different classes a galaxy belongs to during the course of the five epochs. This could be seen as if an individual galaxy can change class during evolution, as is already noticed by van den Bergh (\cite{vdb}) for morphology and Hatton et al. (\cite{galics1}) in GALICS simulated galaxies. However, as discussed in more detail in \cite{paperII}, we find this notion confusing regarding diversification. In astrocladistics, it is preferable to focus on galaxy species: if a galaxy shows a new characteristic typical of a new class, then it should be considered as a new galaxy that somehow have kept some characteristics of its parenthood. Hence these two classes should be close on the tree.

In Fig.~\ref{bestclad}, objects of a given epoch are not always grouped together. Also, two galaxies seem to evolve at different rates, depending also on epoch. This is because evolution does not take place only in time, but also in space. Even if two galaxies are born at the same epoch, they are not formed from the same material (stars, gas and dust), this material not having the same history. In addition these two galaxies do not live in the same environment, and the internal instabilities have no chance to occur at the same time and be of the same intensity. 
Sometimes, two epochs of the same galaxies are reversed on the tree (for instance B1/B2, E4/E5, I1/I4). The explanation should be searched for through an analysis of character evolutions along the cladogram, but this is out of the scope of this paper. In itself this fact is not a problem in view of what has been just said in the previous paragraph. It should be noted that diversity is probably quite low in this sample of galaxies without merger and interaction events.


\section{Conclusions}
\label{conclusions}

Because galaxies are complex objects in evolution, described and characterized by basic constituents (stars, gas and dust), the prerequisites for a phylogenetic analysis using cladistics are satisfied. Using a sample of galaxies resulting from simulations of galaxy formation and evolution, we demonstrate the correctness of our approach. 
Even on a very little diversified galaxy group, like galaxies with no merger and no interaction, cladistics is able to reconstruct the right history from observables. We illustrate the power of such a phylogenetic analysis in providing insights on galaxy physics. For instance, we pinpoint burst characters as being not very pertinent to describe galaxy evolution because they are too variable. 

We are certainly quite far from being able to depict the observed galaxy diversity on a general cladogram. The sample used in this paper is made of galaxies that are too simple as compared to the real world. Other difficulties will certainly arise notably when dealing with interacting objects. The question of the nature and composition of the very first objects can be crucial to the usefulness of astrocladistics because it could multiply the number of different types of ancestors. But conversely, astrocladistics is an excellent tool to investigate this problem, and only a long-range analysis would help. It is important to understand that many developments are possible to improve the analysis shown in this paper: statistical tools can be used to pre-select some groups, many trials could be made concerning character choice,  evolution redundancy and variability, coding of continuous characters should be studied in detail, use of medium and high resolution spectra should be investigated. But only real galaxies deserve as a matter of priority such comprehensive studies.

Cladograms can somehow be seen as a large generalization of the Hubble diagram (Sect.~\ref{outline}). But they have much broader implications and applications. Astrocladistics is a new philosophy for galaxy classification. Beyond paving the way to a new taxonomy, it increases enormously our chances to one day understand galaxy evolution by identifying progenitor classes to today's galaxies back to the very first objects of the Universe.

\begin{acknowledgements}
The authors wish to thank the two referees for very interesting and helpful comments on the manuscript. This work uses the GalICS/MoMaF Database of Galaxies (http://galics.iap.fr). We warmly thank Bruno Guiderdoni and Jeremy Blaizot from the GALICS project for invaluable help and discussions. All the GALICS team should be thanked for providing such a useful database. Emmanuel Davoust kindly read the manuscript and helped us much in improving it. Most computations of the results presented in this paper were performed under the CIMENT project in Grenoble. The contribution of EJPD is publication 2006-003 of the Institut des Sciences de l'\'Evolution de Montpellier (UMR 5554 - CNRS).
\end{acknowledgements}

\begin{table}
	\begin{center}
		\begin{tabular}{ll}
A1& 5522317054017003037100567777202200000000000007777777777777777777777733333301113664633456211 \\
A2& 5443527044017105047200567776012211111111411117777777777777777777777755555502224665643546222 \\
A3& 4356747133027106057200445543022222222222632227777777777777777777777756666602344666644536222 \\
A4& 2007156100027107067300555443052245445444755557777777777777777777777766666602555666644536322 \\
A5& 1007056100027107067300665543073377777777777777777777777777777777777766666612665666644536322 \\
B1& 1120232111000000000000222111211112111111621117777777777777777777777777777777777777777777777 \\
B2& 1141453211003000013100444443122122222222622227777777777777777777777733344466116666666656666 \\
B3& 0002143200774700014100444433142144444444744440000000600656666666366677777777777777777777777 \\
B4& 0002043200774700014100777765263366666666666664556663345000000001003377777777777777777777777 \\
B5& 0002043200774700014100665543277777777777777776666656066563355630650077777777777777777777777 \\
C1& 0000001000000000000000112221601000000000200007777777777777777777777777777777777777777777777 \\
C2& 0000021100023370763677222221311111221112621117777777777777777777777722222250000000000000000 \\
C3& 0000122100033300073700455543222222222222622227777777777777777777777755555541223665633446211 \\
C4& 0001021100033300073700555443252255555555755557777777777777777777777766666642555666655636322 \\
C5& 0001021100033300073700665543377777777777777777777777777777777777777766666652665666655636322 \\
D1& 7772727077027104047200445554012101001110411117777777777777777777777732222201113665633466211 \\
D2& 4124456111027105067200433332032133333333733337777777777777777777777755555602234665643546222 \\
D3& 1004057100027105067200555443062256556555766667777777777777777777777755666612454665644536222 \\
D4& 1004057100027105067200665543075477777777777777777777777777777777777766666612664666644536222 \\
D5& 0004057100027105067200665543177777777777777777777777777777777777777766666622664666644536222 \\
E1& 3341425033015101045200556665112201000000311117777777777777777777777732233222114666644566222 \\
E2& 1002134100015101045200677776133233333333533337777777777777777777777755556622235666654646322 \\
E3& 0002034100015101045200555443162266666666766667777777777777777777777755666632455666654636322 \\
E4& 0002034100015101045200665543176677777777777777777777777777777777777766666632565666654636322 \\
E5& 0002034100015101045200766544277777777777777777777777777777777777777766666642665666654636322 \\
F1& 3331215033016001026100556666112201000000311117777777777777777777777721111121113664633466211 \\
F2& 3342536033026103056300556665112211111111411117777777777777777777777744555512224666644546322 \\
F3& 2002007000037107077300544443132233333333633337777777777777777777777755555502235666654646322 \\
F4& 0001007000037107077300665543175677777777777777777777777777777777777766666612555666654636322 \\
F5& 0001007000037107077300765543277777777777777777777777777777777777777766666622665666654636322 \\
G1& 0000031200000000000000211110321122222222722227777777777777777777777777777777777777777777777 \\
G2& 0000031200000000000000443332352155555555755557777777777777777777777777777777777777777777777 \\
G3& 0000031200000000000000554433362277777777777667777777777777777777777777777777777777777777777 \\
G4& 0000031200000000000000665443476677777777777777777777777777777777777777777777777777777777777 \\
G5& 0000031200000000000000665543477777777777777777777777777777777777777777777777777777777777777 \\
H1& 0010031200000000000000000000410011111111511117777777777777777777777777777777777777777777777 \\
H2& 0000020200000000000000332221331134444334743337777777777777777777777777777777777777777777777 \\
H3& 0000020200000000000000544433362266666666766667777777777777777777777777777777777777777777777 \\
H4& 0000020200000000000000655443476677777777777777777777777777777777777777777777777777777777777 \\
H5& 0000020200000000000000665543576777777777777777777777777777777777777777777777777777777777777 \\
I1& 0010002011003100023222111111400000000000300007777777777777777777777700000050000100000041000 \\
I2& 0000003010013100033300445554301100000000300007777777777777777777777722222241112653622366111 \\
I3& 0000122100013100033300434332221123332222732227777777777777777777777755555542335666654646322 \\
I4& 0011232100023201053500444433221122222222622227777777777777777777777755556642225666655646322 \\
I5& 0011242200023301063600444443231133333333733337777777777777777777777766666642666666665636322 \\
J1& 0000070700000000000000000000720023333223732227777777777777777777777777777777777777777777777 \\
J2& 0010141200000000000000222211321122222222722227777777777777777777777777777777777777777777777 \\
J3& 0000041300000000000000444433252256556555765557777777777777777777777777777777777777777777777 \\
J4& 0000041300000000000000555443372277777777777777777777777777777777777777777777777777777777777 \\
J5& 0000041300000000000000665543476777777777777777777777777777777777777777777777777777777777777 \\
		\end{tabular}
	\end{center}

	\caption{Coded matrix with photometry values taken as colors with respect to the K-band for each component. Column numbers correspond to character numbers listed in Table~\ref{tabchar}. This table is available on http://hal.ccsd.cnrs.fr/aut/fraix-burnet.}
	\label{charcod}
\end{table}

\end{document}